\documentclass[aps,prl,twocolumn,amsmath,amssymb,amsfonts,nofootinbib,long,floatfix]{revtex4}
\usepackage{epsfig,latexsym,bm}

\begin{document}

\title{Creation of Universes from the Third-Quantized Vacuum}

\author{Leonardo Campanelli$^{1}$}
\email{leonardo.s.campanelli@gmail.com}
\affiliation{$^1$All Saints University, Asudom Academy of Science, 5145 Steeles Ave., Toronto (ON), Canada}

\date{\today}


\begin{abstract}
We calculate the average numbers of closed, flat, and open universes spontaneously created from
nothing in third quantization. The creation of universes is exponentially suppressed
for large values of the kinetic energy of the inflaton, while for small kinetic energies
it is exponentially favoured for closed universes over flat and open ones: For a scale
of inflation less than about $2 \times 10^{16}$GeV, the ratio of the number of closed universes
to either the number of flat or open universes is
\begin{equation}
\frac{n_{closed}}{n_{flat,open}} \gtrsim 10^{10^{10}} . \nonumber
\end{equation}
\end{abstract}



\maketitle
\newpage


\section{I. Introduction}

In their seminal paper~\cite{Hosoya-Morikawa}, Hosoya and Morikawa explored the consequences of
the quantization of the wave function of the Universe, now known as third quantization.
The main motivation was to overcome the problem of the probabilistic interpretation
of the wave function of the Universe, solution of the Wheeler-DeWitt
equation: since the latter is a hyperbolic second-order differential equation,
it does not admit conserved quantities that are positive definite.
Their proposal of a quantum field theory of the Universe resembles to the
one that successfully solved the problem of negative probability
in the case of the Klein-Gordon equation.

As a consequence of their investigation, Hosoya and Morikawa
discovered that universes are spontaneously created from ``nothing''
(the third-quantized vacuum), in the same way particles can be created from vacuum if
the external potential is time dependent. In third-quantization, the time-dependent potential
(the Wheeler-DeWitt potential) naturally arises from Einstein gravity, and
the time variable is played by the (logarithm) of the expansion parameter.

In their paper, Hosoya and Morikawa calculated the average number of flat universes created
from nothing in the presence of an homogeneous scalar field (the inflaton).
Recently enough, Kim~\cite{Kim} calculated the number of closed and open universes in the case
of vanishing potential of the scalar field. The aim of this paper is to
evaluate this number in the general case of nonvanishing scalar potential.

The plan of the paper is as follows. In Sec. II, we briefly review third quantization
in minisuperspace and in particular the mechanism of creation of universes from nothing.
In Sec. III, an analogy between universe creation and quantum potential scattering is analyzed.
This analogy will allow us to use standard WKB methods used in quantum mechanics for the
calculation of the number of cerated universes.
In Sections IV, V, and VI, we calculate the average numbers of flat, closed, and open
universes created out of nothing in the particular case of constant scalar potential,
both using WKB approximation and an approximate form of the Wheeler-DeWitt potential.
In Sec. VII, we discuss our result and we draw our conclusions.

\section{II. Third quantization and the creation of universes from nothing}

\subsection{IIa. Nothingness and multiverse}

The Wheeler-DeWitt equation in homogeneous
and isotropic minisuperspace is (using the units $\hbar = c = 4\pi G/3 = 1$)
\begin{equation}
\label{motion}
\left[\frac{\partial^2}{\partial \alpha^2} - \frac{\partial^2}{\partial \phi^2}
+ U(\alpha,\phi) \right] \! \Psi(\alpha,\phi) = 0,
\end{equation}
where $\alpha = \ln a$, with $a$ being the expansion parameter, $\phi$ is a real
scalar field (which we identify as the inflaton), $k$ is the signature of the spatial curvature,
and
\begin{equation}
\label{U}
U(\alpha,\phi) = V^2 e^{4\alpha} \left[2V(\phi) e^{2\alpha} - k \right]
\end{equation}
%
%
is the Wheeler-DeWitt potential.
The spatial volume $V$ is equal to $2\pi^2$ for closed universes ($k=1$).
For flat ($k=0$) and open ($k=-1$) universes, $V$ is a normalization volume
that can be taken as the (finite) volume of the region under consideration.

In third quantization, the ``Universe field'' $\Psi(\alpha,\phi)$ is
expanded in normal modes with the coefficients of expansions being the
annihilation and creation operators. A Fock space can be constructed
starting from a vacuum state which represents a state of {\it nothing},
a state in which even space-time does not exist.

Following~\cite{Hosoya-Morikawa}, we assume that the Universe is a neutral scalar.
In this case, we can write the Universe wave function in the in-Fock space as
\begin{equation}
\label{A}
\Psi(\alpha,\phi) =
\int \!\! \frac{dp}{2\pi} \! \left(c_p \, \psi_p(\alpha) \, e^{ip \phi} + c_p^{\dag} \, \psi_p^*(\alpha) \, e^{-ip \phi} \right) \!,
\end{equation}
where the subscript $p$ labels the mode function and its physical
meaning will be discussed later. Here, the annihilation and creation operators $c_p$ and $c_p^{\dag}$
satisfy the usual commutation relations, $[c_p, c_{p'}^{\dag}] = 2\pi \delta_{p p'}$ and
$[c_p, c_{p'}] = [c_p^{\dag}, c_{p'}^{\dag}] = 0$.
The functions $\psi_p(\alpha)$ are positive frequency
solutions (with respect to $\alpha$) of the Schrodinger-like equation
\begin{equation}
\label{motion3}
\ddot{\psi}_p = U_p \psi_p,
\end{equation}
where a dot indicates a derivative with respect to $\alpha$, and
\begin{equation}
\label{Up}
U_p(\alpha) = -p^2 - V^2 e^{4\alpha} \left(2V_0 e^{2\alpha} - k \right) \!.
\end{equation}
Hereafter, we consider only the case of a constant
scalar potential $V(\phi) = V_0$. Also, we assume $V_0 \neq 0$ throughout the paper, with
the expection of Section VIa, where we discuss the case of open universes with
vanishing scalar potential.

In order to have a self-consistent quantization, the mode $\psi_p(\alpha)$
must satisfy the Wronskian condition
$\psi_p \dot{\psi}_p^{*} - \dot{\psi}_p \psi_p^{*} = i$.

The vacuum state $|0\rangle$ is defined by
\begin{equation}
\label{multiverse}
\forall p \in \mathbb{R} \!: ~ c_p |0\rangle = 0 ~~ \mbox{(nothinghness)}
\end{equation}
%
and is normalized as $\langle 0|0\rangle = 1$.
The state $c_{p}^{\dag} |0\rangle$ represents the single universe,
the state $c_{p}^{\dag} c_{p'}^{\dag}|0\rangle$ represents a
double universe and, in general, the state
\begin{equation}
\label{multiverse}
\prod_{i=1}^{N} c_{p_i}^{\dag} |0\rangle  ~~ \mbox{(multiverse)}
\end{equation}
%
represents the {\it multiverse}, namely a state with $N$ universes each
of them labeled by $p_i$.

\subsection{IIb. Universes from nothing}

As in the case of quantum field theory
in curved spacetime, the vacuum state is not unique. Different inequivalent
physical vacua can be introduced in different region in minisuperspace.
In particular, we can define in- and out-regions for $\alpha \rightarrow -\infty$
and  $\alpha \rightarrow +\infty$, respectively, to which there
corresponds in- and out-vacuum states.


The in-vacuum state contains no in-universes in the in-region.
Such a ``Bunch-Davies vacuum'' can be constructed by
solving the Wheeler-DeWitt equation for the Universe states and then
by fixing the constants of integrations appearing in the
general solution by matching the latter with the corresponding
adiabatic solution for $\alpha \rightarrow -\infty$. Accordingly,
we can construct the in-Fock space based on the in-vacuum by
repeatedly applying the in-creation operator on the in-vacuum state.
Another Fock state can be constructed in this way, but this time
starting from a out-region $\alpha \rightarrow +\infty$.

It is clear from the above discussion that the two Fock spaces based on the two different
choices of the (Bunch-Davies) vacuum state are both physically
acceptable and must be then related. In particular, there will be a relation
between the in- and out-modes $\psi_p^{({\rm in})}$ and $\psi_p^{({\rm out})}$,
as well as a relation between the in- and out-creation and annihilation operators.
In order to find these relations, let us observe that
if $\psi_p^{(1)}$ and $\psi_p^{(2)}$ are two solutions of Eq.~(\ref{motion3}),
the following inner product is conserved,
\footnote{The possibility of introducing an inner product
remains valid even in superspace due to hyperbolicity of the Wheeler-DeWitt equation
(see, e.g.,~\cite{Kim} and references therein).}
\begin{equation}
\label{inner}
\langle \psi_p^{(1)} | \psi_p^{(2)} \rangle =
-i (\psi_p^{(1)} \dot{\psi}_p^{(2)} - \dot{\psi}_p^{(1)} \psi_p^{(2)}).
\end{equation}
We can then introduce the time-independent quantities
\begin{eqnarray}
\label{alphabeta1}
&& \alpha_p = \langle \psi_p^{({\rm in})} | \psi_p^{({\rm out}) *} \rangle, \\
\label{alphabeta2}
&& \beta_p = -\langle \psi_p^{({\rm in})} | \psi_p^{({\rm out})} \rangle,
\end{eqnarray}
and expand the in-$\psi$ mode in terms of the out-$\psi$ mode as
\begin{equation}
\label{in-out}
\psi_p^{({\rm in})} =
\alpha_p \psi_p^{({\rm out})} + \beta_p \psi_p^{({\rm out}) *},
\end{equation}
where we used the fact that
\begin{equation}
\label{inner2} \langle \psi_p^{({\rm in})} | \psi_p^{({\rm in}) *} \rangle =
\langle \psi_p^{({\rm out})} | \psi_p^{({\rm out}) *} \rangle = 1.
\end{equation}
Equation~(\ref{in-out}) is the wanted relation between the $\psi$ in-
and out- modes. A relation of this type is know as Bogolubov
transformation and the quantities $\alpha_p$ and $\beta_p$
are called Bogolubov coefficients. They satisfy the relation
\begin{equation}
\label{relation} |\alpha_p|^2 - |\beta_p|^2  = 1,
\end{equation}
which can be easily derived from their defining equations.
To find the relation between the in- and out-creation and annihilation operators,
we insert the Bogolubov transformation in Eq.~(\ref{A})
and compare the result with the the expression of $\Psi$ defined in the out-Fock space.
 We find
$c_p^{({\rm out})} = \alpha_p \, c_p^{({\rm in})} - \beta_p^* \, c_{-p}^{({\rm in}) \dag}$.
From the above equation,
it follows immediately that the two Fock spaces
based on the two choices $|0, {\rm in} \rangle$ and $|0, {\rm out} \rangle$ of
the vacuum are generally different.
In particular, the in-vacuum state will contain out-universes as long as $\beta_p \neq 0$,
\begin{equation}
\label{n}
n_p = \langle 0, {\rm in} |N^{({\rm out})}_p| 0, {\rm in} \rangle = |\beta_p|^2,
\end{equation}
where $N^{({\rm out})}_p = c_p^{({\rm out}) \dag} c_p^{({\rm out})}$
is the number operator in the out-Fock space.
Note that universes are created in pairs with opposite $p$. 

\subsection{IIc. Labeling universes}


\begin{figure*}[t!]
\begin{center}
\includegraphics[clip,width=0.6\textwidth]{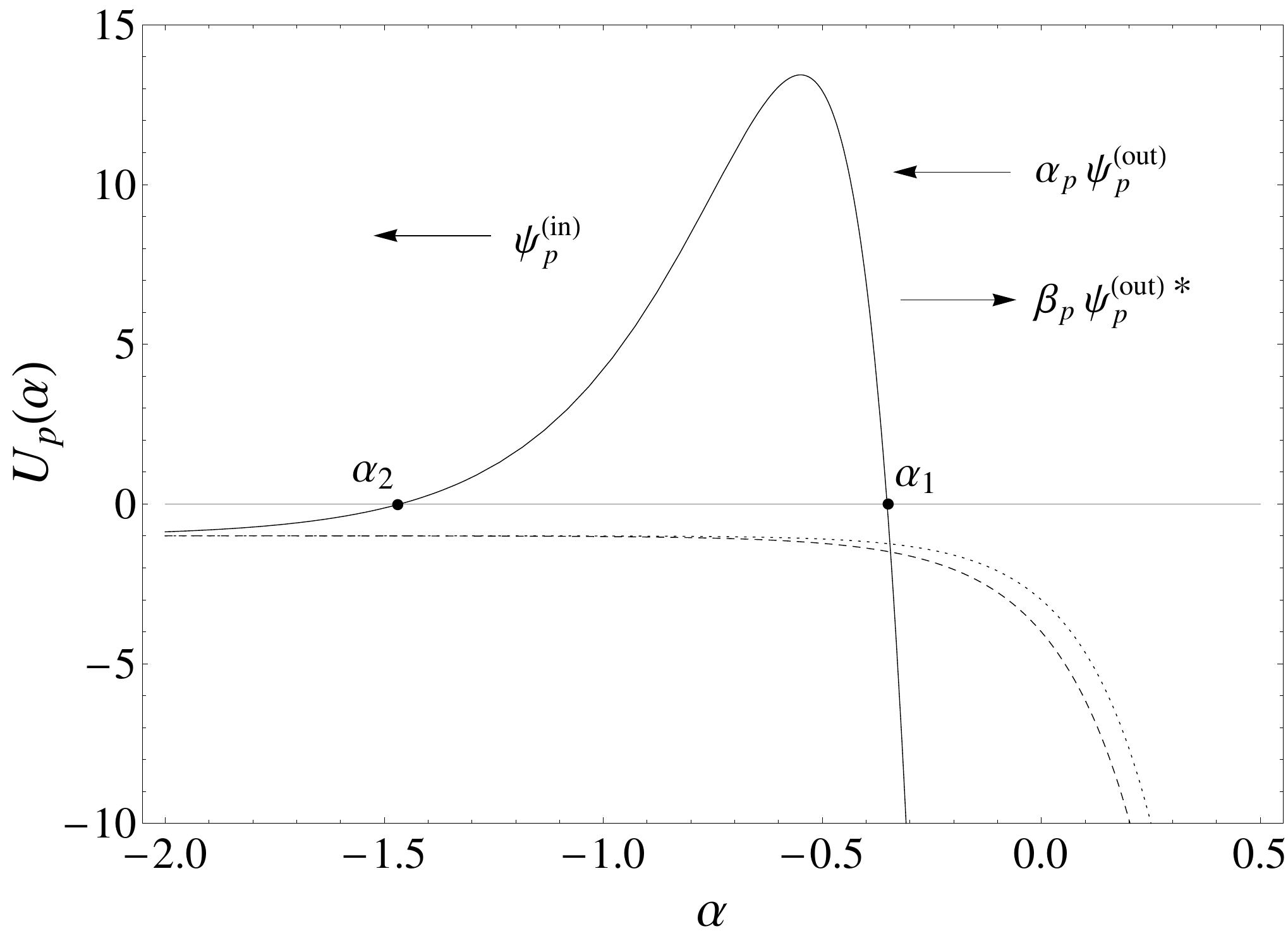}
\caption{The Wheeler-DeWitt potential $U_p(\alpha)$ in Eq.~(\ref{Up}) for $p = V_0 = 1$
and $k=1$ (continuous line), $k=-1$ (dashed line), and $k=0$ (dotted line). For
open and flat universes, the volume $V$ has been taken equal to unity.
The points $\alpha_1$ and $\alpha_2$ are the classical turning points in the WKB picture
discussed in Section III.}
\end{center}
\end{figure*}


Classically, the canonical momentum conjugate
to $\phi$ is given by $p_\phi^{({\rm cl})} = Va^3  d\phi/dt$~\cite{Hosoya-Morikawa}.
Accordingly, $p$ is related to the kinetic energy (density) of the scalar field,
$K_\phi$, through
\begin{equation}
\label{KK}
K_\phi = \frac12 \! \left( \! \frac{d\phi}{dt} \right)^{\!2} = \frac{p^2}{2Va^3} \, .
\end{equation}
Thus, $p$ essentially labels universes with different amounts of
kinetic energy of the inflaton.
In the out-region, where the created universes behave classically,
the expansion is governed by the usual Friedmann equation
\begin{equation}
\label{Friedmann}
H^2 = \left( \! \frac{d\phi}{dt} \right)^{\!2} +2V(\phi) -\frac{k}{a^2} \, ,
\end{equation}
where $H = (da/dt)/a$ is the Hubble parameter.
Taking into account Eq.~(\ref{KK}), the Friedmann equation takes the form
\begin{equation}
\label{Friedmann2}
H^2 = \frac{p^2}{V^2 a^6} + 2V_0 - \frac{k}{a^2} = -\frac{U_p(\alpha)}{V^2e^{6\alpha}} \, ,
\end{equation}
where $U_p(\alpha)$ is given by Eq.~(\ref{Up}) for the case of constant scalar potential
(see Fig.~1).

{\it Flat universes.} -- For flat universes, the solution of Eq.~(\ref{Friedmann2})
with $a(0)=0$ is easily found,
\begin{equation}
\label{P1}
a(t) = \left(\frac{|p|t}{V}\right)^{\!1/3} \! \sinh(3t/a_{\rm cr}),
\end{equation}
where we have defined
\begin{equation}
\label{acr}
a_{\rm cr} = 1/\sqrt{2V_0}.
\end{equation}
The above expression for the expansion parameter is well approximated by
\begin{eqnarray}
\label{P2}
a(t) \!\!& \simeq &\!\!
\left\{ \begin{array}{lll}
  \!\! \left(\frac{|p|t}{V}\right)^{\!1/3} \!,                    & ~ a \lesssim a_{\rm cr} r^{1/6}, \\
\\
  \!\! \left(\frac{|p|a_{\rm cr}}{V}\right)^{\!1/3} e^{t/a_{\rm cr}-1},   & ~ a \gtrsim a_{\rm cr} r^{1/6},
  \end{array}
  \right.
\end{eqnarray}
where
\begin{equation}
\label{r}
r = \left(\!\frac{p}{Va_{\rm cr}^2}\!\right)^{\!2} \! .  
\end{equation}
Thus, flat universes created in the out region with sufficiently large expansion parameter,
$a \gtrsim a_{\rm cr} r^{1/6}$, undergo inflation, $a(t) \propto e^{\sqrt{2V_0} t}$, while
those created with small expansion parameter, $a \lesssim a_{\rm cr} r^{1/6}$,
are dominated by the kinetic energy of the scalar field, $a(t) \propto t^{1/3}$, and do not inflate
(see the upper panel of Fig.~2).

{\it Closed universes.} --  Closed universes are created in the out region only if $\alpha > \alpha_1$
(corresponding to $H^2 > 0$), where $\alpha_1$ is the largest zero of the potential
$U_p(\alpha)$ (see Fig.~1). This corresponds to expansion parameters
$a > \sqrt{x_1} a_{\rm cr}$, where $x_1$ is defined in Eq.~(\ref{x1}).
If $a < \sqrt{x_1} a_{\rm cr}$, the square of the Hubble parameter is negative,
which indicates a recollapsing universe. This analysis is true when $r < 4/27$ (see discussion
in Section V), while for $r > 4/27$ universes with any value of $a$ can be created.
Using the results of Section V (see in particular Fig.~3), the root $x_1(r)$ is
in the interval $2/3 < x_1(r) < 1$ for $0< r < 4/27$. For the sake of simplicity
and convenience, let us assume that created universes recollapse when $a \lesssim a_{\rm cr}$ for
$r \lesssim 1$.
Assuming that either $a \gtrsim a_{\rm cr}$ or $r \gtrsim 1$, the expression for the
expansion parameter can be approximated as
\begin{eqnarray}
\label{P3}
a(t) \!\!& \simeq &\!\!
\left\{ \begin{array}{lll}
  \!\! \left(\frac{|p|t}{V}\right)^{\!1/3} \!,                     & ~ a \lesssim a_{\rm cr} r^{1/6}, ~ r \gtrsim 1, \\
\\
  \!\! \left(\frac{|p|a_{\rm cr}}{V}\right)^{\!1/3} e^{t/a_{\rm cr}-1},    & ~ a \gtrsim a_{\rm cr} \max[1,r^{1/6}].
  \end{array}
  \right.
\end{eqnarray}
The upper branch of $a(t)$ in the above equation corresponds to the case of a dominant kinetic term
in the Hubble parameter, while the lower branch to the case of a dominant potential term.
In the former case, universes inflate, in the latter they do not (see the middle panel of Fig.~2).

{\it Open universes.} --  For open universes, the approximated solution of Eq.~(\ref{Friedmann2}) reads
\begin{eqnarray}
\label{P4}
a(t) \!\!& \simeq &\!\!
\left\{ \begin{array}{lll}
  \!\! \left(\frac{|p|t}{V}\right)^{\!1/3} \!,        & ~ a \lesssim a_{\rm cr} r^{1/4}, ~ r \lesssim 1, ~ \mbox{or} \\
                                                      & ~ a \lesssim a_{\rm cr} r^{1/6}, ~ r \gtrsim 1,  \\
\\
  \!\! \left(\frac{|p|a_{\rm cr}}{V}\right)^{\!1/3} e^{t/a_{\rm cr}-1}, & ~ a \gtrsim a_{\rm cr}, ~ r \lesssim 1,
                                                                        ~\mbox{or} \\
                                                      & ~ a \gtrsim a_{\rm cr} r^{1/6}, ~ r \gtrsim 1,  \\
\\
  \!\! t,                                             & ~ a_{\rm cr} r^{1/4} \lesssim a \lesssim a_{\rm cr}, ~ r \lesssim 1.
  \end{array}
  \right.
\end{eqnarray}
The three branches correspond to a dominant kinetic, potential, and curvature term in
the Hubble parameter. The lower panel of Fig.~2 graphically shows the case of open
universes.


\begin{figure}[t!]
\begin{center}
\includegraphics[clip,width=0.48\textwidth]{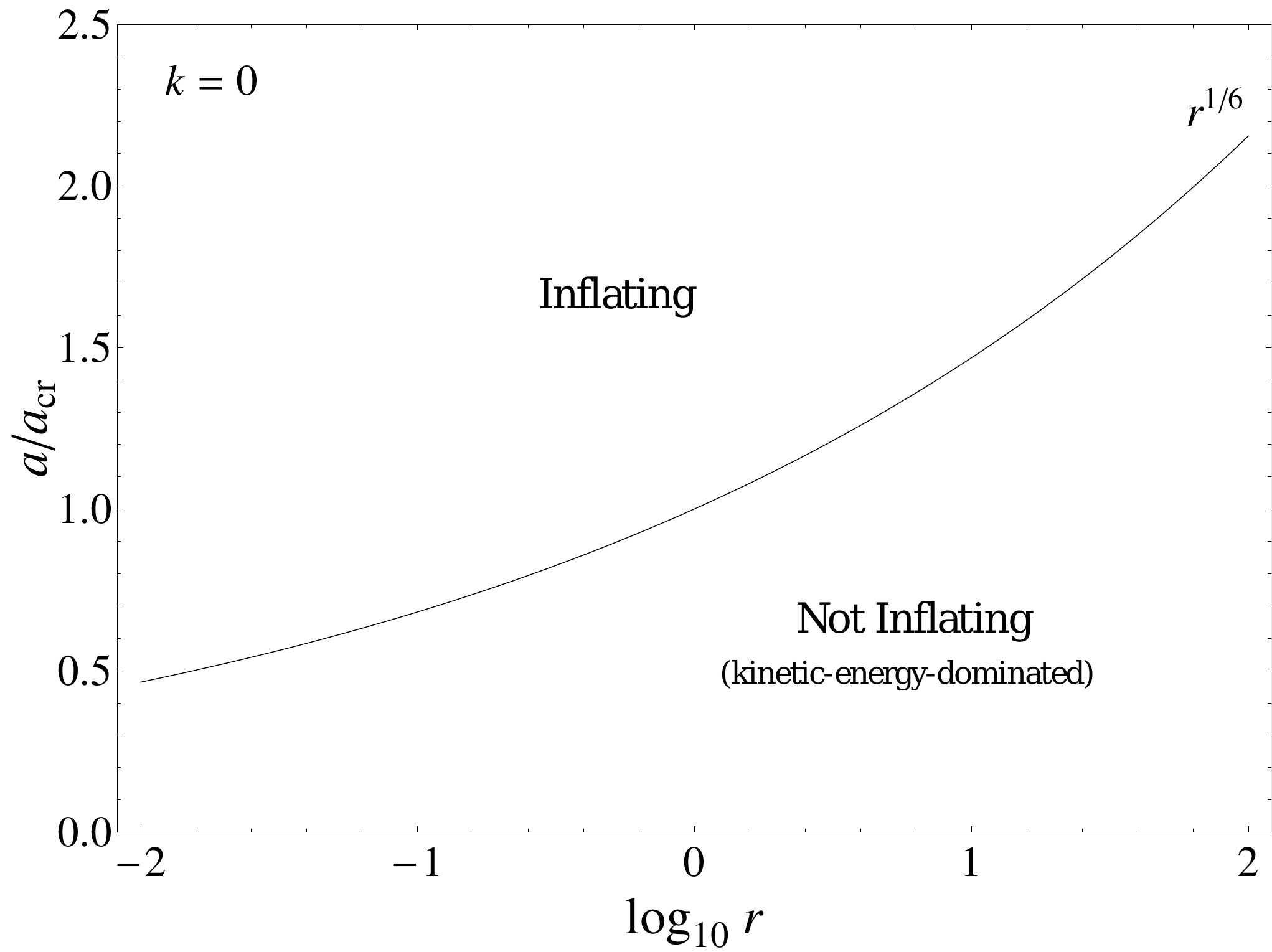}

\hspace{0.3cm}

\includegraphics[clip,width=0.48\textwidth]{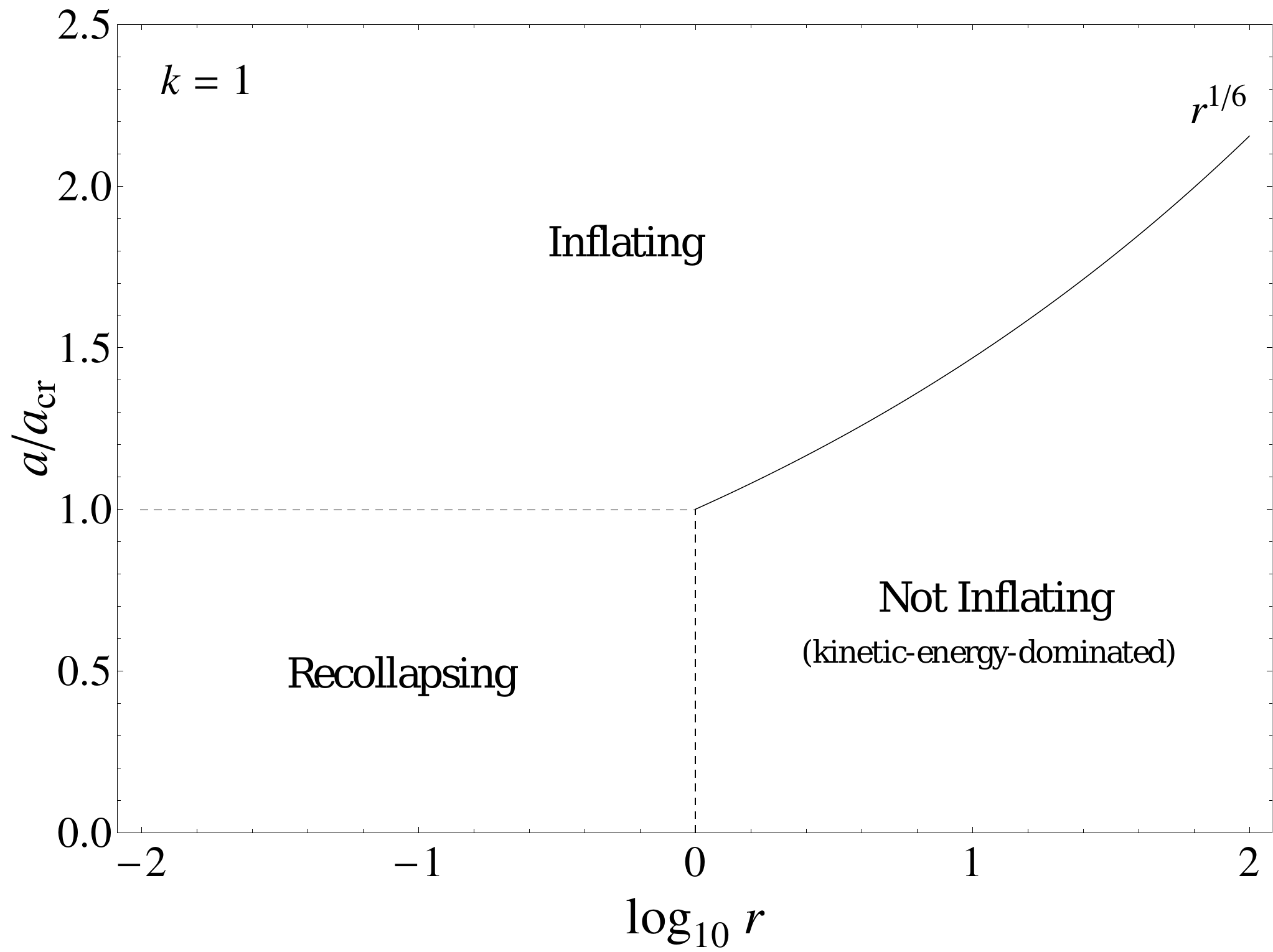}

\vspace{0.3cm}

\includegraphics[clip,width=0.48\textwidth]{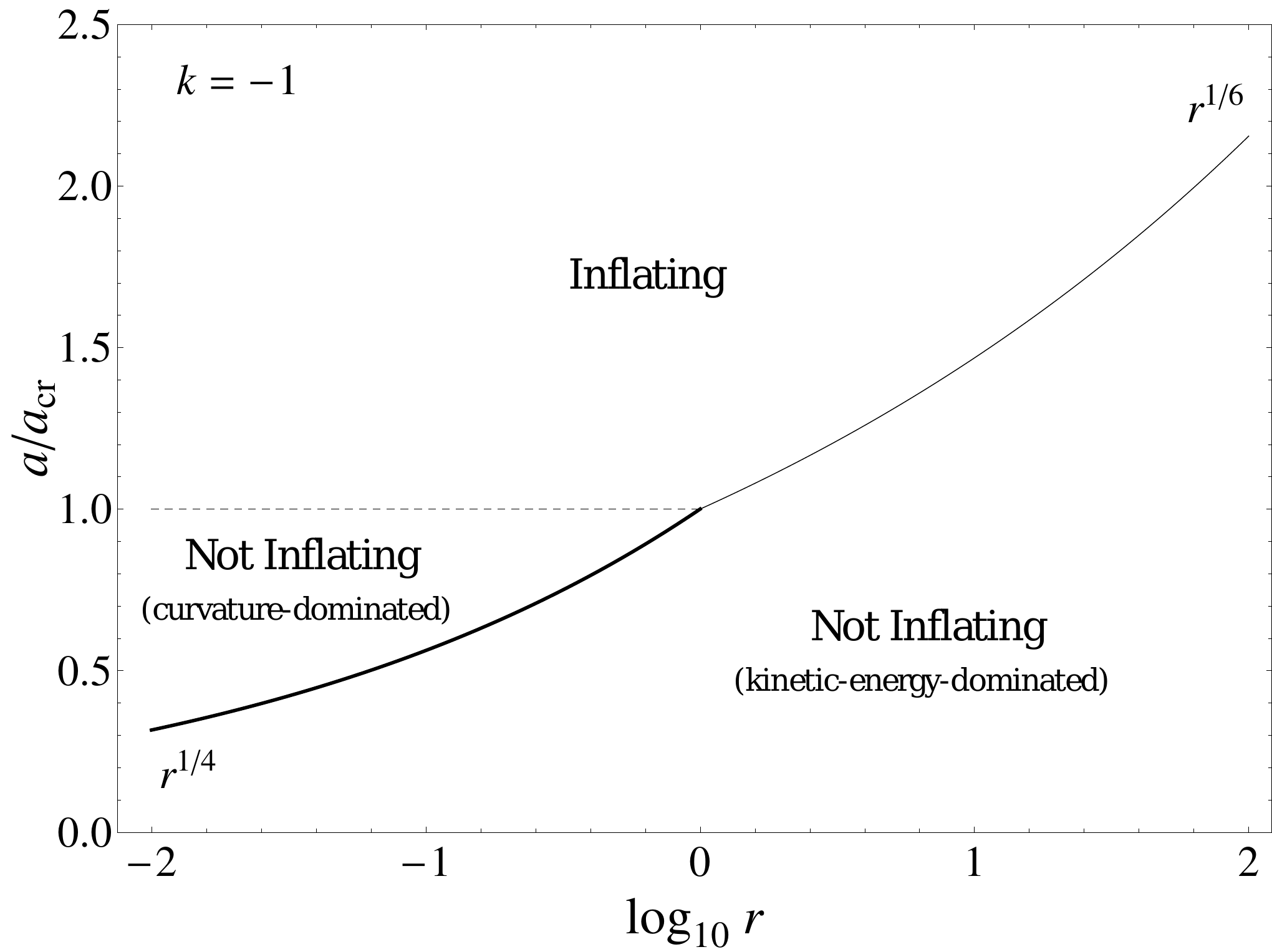}
\caption{The expansion parameter of flat (upper panel), closed (middle panel),
and open (middle panel) universes created in the out region as a function of the parameter
$r$ defined by Eq.~(\ref{r}).}
\end{center}
\end{figure}


\section{III. Universe creation analogy with quantum potential scattering}

\subsection{IIIa. General considerations}

The Wheeler-DeWitt equation~(\ref{motion3})
for the $\psi$ modes is formally equal to the one-dimensional Schrodinger
equation with zero energy, mass equal to $1/2$, and potential energy
$U_p(\alpha)$, with $\alpha$ taking the place of the spatial coordinate
and $p$ being an external parameter.
Continuing the analogy, Eq.~(\ref{in-out}) connecting the
$\psi_{p}^{(\rm in)}$ and $\psi_{p}^{(\rm out)}$ modes describes the
scattering of $\psi$-waves off the potential $U_p$, the incident,
reflected, and transmitted waves being
\begin{eqnarray}
\label{waves}
&& \psi_p^{({\rm inc})} = \alpha_p \psi_p^{(\rm out)}, \\
&& \psi_p^{({\rm ref})} = \beta_p  \psi_p^{(\rm out) *}, \\
&& \psi_p^{({\rm tr})}  = \alpha_p \psi_p^{(\rm in)},
\end{eqnarray}
respectively, as illustrated in Fig.~1.
Moreover, one can define a density current associated to any $\psi$-mode as
\begin{equation}
\label{current}
j_p = \langle \psi_p | \psi_p^* \rangle.
\end{equation}
The conservation of the current~(\ref{current}), $\dot{j}_p = 0$, follows directly from the conservation of the inner product. The incident, reflected, and transmitted currents are then
\begin{eqnarray}
\label{currents}
&& j_p^{({\rm inc})} =
  \langle \alpha_p \psi_p^{({\rm out})} | \alpha_p^* \psi_p^{({\rm out}) *} \rangle = |\alpha_p|^2, \\
&& j_p^{({\rm ref})} =
  \langle \beta_p \psi_p^{({\rm out}) *} | \beta_p^* \psi_p^{({\rm out})} \rangle = -|\beta_p|^2, \\
&& j_p^{({\rm tr})} =
  \langle \psi_p^{({\rm in})} | \psi_p^{({\rm in}) *} \rangle = 1,
\end{eqnarray}
where we used Eqs.~(\ref{inner2}).

Taking into account Eq.~(\ref{n}) and the Bogoliubov condition~(\ref{relation}), we find
the reflection and transmission coefficients
\begin{eqnarray}
\label{Ref}
&& R_{p} = -\frac{j_{p}^{({\rm ref})}}{j_{p}^{({\rm inc})}} = \frac{n_{p}}{1+n_{p}} \, , \\
\label{Tr}
&& T_{p} = \frac{j_{p}^{({\rm tr})}}{j_{p}^{({\rm inc})}} = \frac{1}{1+n_{p}} \, ,
\end{eqnarray}
from which the unitarity condition $R_{p} + T_{p} = 1$ directly follows.

It is clear that if $p^2 < \max U_0(\alpha)$, then the ``particle''
described by the wave function $\alpha_p \psi_p^{(\rm out)}$ will
penetrate through the potential barrier $U_p(\alpha)$.
To ``particles'' which deeply penetrate into the barrier, $p^2 \ll \max U_0(\alpha)$,
there will correspond a large reflection coefficient and, in turn,
by Eq.~(\ref{Ref}), a large ``particle'' number $n_p$.
On the other hand, if $p^2 > \max U_0(\alpha)$, the
``particle'' is reflected above the barrier. For $p^2 \gg \max U_0(\alpha)$,
the reflection coefficient for scattering above the barrier will be small.
To this case, there will correspond a small production of ``particles'', $n_p \ll 1$.

\subsection{IIIb. WKB approximation}

The usefulness of Eqs.~(\ref{Ref}) and (\ref{Tr}) resides in the fact that if the
potential $U_p$ is slowly varying, in the sense specified below,
one can apply the standard semiclassical (WKB) results for the
reflection and transmission coefficients. Using the formal equivalence
of the two problems of potential scattering in
quantum mechanics and the creation of
universes out from the vacuum in third quantization, one can then
find the expression for the universe number $n_p$.
The WKB approximation is valid whenever
the potential $U_p$ satisfies the semiclassical condition~\cite{Landau}
\begin{equation}
\label{conditionU}
\left|\frac{\dot{U}_p}{2U_p^{3/2}} \right| \ll 1.
\end{equation}
It can be verified that the above condition is satisfied
for the Wheeler-DeWitt potential~(\ref{U}) for values of $\alpha$ far from the
turning points, where the WKB approximation is in general not valid.

{\it Large universe number.} -- Let us consider the case of closed universes, $k>1$ (see Fig.~1).
Accordingly, there will be two classical turning points, $\alpha_2(p) < \alpha_1(p)$,
for a deep penetration through the potential barrier.
Since in this case $R_p \simeq 1$, and then $T_p \ll 1$, we have from Eq.~(\ref{Tr}),
$n_p \simeq T_p^{-1} \gg 1$. Using the standard result for the
expression of the transmission coefficient in WKB approximation~\cite{Landau},
we find
\begin{equation}
\label{sm2}
n_p = e^{2S_p},
\end{equation}
where
\begin{equation}
\label{sm3}
S_p = \int_{\alpha_2(p)}^{\alpha_1(p)} \! d\alpha \sqrt{U_p(\alpha)} \, .
\end{equation}

{\it Small universe number.} -- The probability that a ``particle'' is scattered above the
potential barrier is small for large values of $p^2$ compared
to the height of the Wheeler-DeWitt barrier $U_0(\alpha)$. This is true for
closed, flat, and open universes. Using Eq.~(\ref{Ref}), we
then have $n_p \simeq R_p \ll 1$. Using the standard result for the
expression of the reflection coefficient in WKB approximation~\cite{Landau},
we find
\begin{equation}
\label{sm4}
n_p = e^{-4 \mathrm{Im} \sigma_p},
\end{equation}
where
\begin{equation}
\label{sm5}
\sigma_p = \int_{\alpha_R}^{\alpha_I(p)} \! d\alpha \sqrt{-U_p(\alpha)} \, .
\end{equation}
Here, $\alpha_I(p)$ is the so-called imaginary turning point,
the complex solution of the equation $U_p(\alpha) = 0$ for $p^2 > \max U_0(\alpha)$,
and $\alpha_R$ is an arbitrary and inessential real parameter.
The integration in Eq.~(\ref{sm5}) has to be performed in the complex upper half-plane,
$\mbox{Im}[\alpha_I(p)] > 0$.
If the equation for the imaginary turning point admits more than one solution,
one must select the one for which $\sigma_p$ is smallest~\cite{Landau}.

\section{IV. Creation of flat inflationary universes}

{\it Exact solution.} -- The case $k=0$ was analyzed by Hosoya and Morikawa~\cite{Hosoya-Morikawa}.
An exact solution for the number of created universes is given by
\begin{equation}
\label{HM}
n_p = \frac{1}{e^{2\pi |p|/3}-1}
\end{equation}
and is, interestingly enough, independent on $V_0$. For large $|p|$, $n_p $ is exponentially suppressed, while for small $|p|$, $n_p$ is inversely proportional to $|p|$.
Equation~(\ref{HM}) is easily found by inserting the Bunch-Davies, in- and out-solutions
of Eq.~(\ref{motion3}) (with $k=0$),
\begin{eqnarray}
\label{hp}
&& \!\!\!\!\!\!\!\!\!\!\!\!\!\!\!\!\!\!
   \psi_p^{(\rm in)} = \sqrt{\frac{\pi}{6}} \,
                               \sinh^{-1/2}(p\pi/3) J_{-ip/3}(V\sqrt{2V_0}e^{3\alpha}/3), \\
\label{hp2}
&& \!\!\!\!\!\!\!\!\!\!\!\!\!\!\!\!\!\!
    \psi_p^{(\rm out)} = \sqrt{\frac{\pi}{12}} \,
                                  e^{-p\pi/6} H_{-ip/3}^{(2)}(V\sqrt{2V_0}e^{3\alpha}/3),
\end{eqnarray}
into Eq.~(\ref{alphabeta2}) and then using Eq.~(\ref{n}).
Here, $J_\nu (x)$ is the Bessel function of first kind and $H_\nu^{(2)}(x)$ is the Hankel function of
second kind~\cite{Abramowitz}.


\begin{figure*}[t!]
\begin{center}
\includegraphics[clip,width=0.48\textwidth]{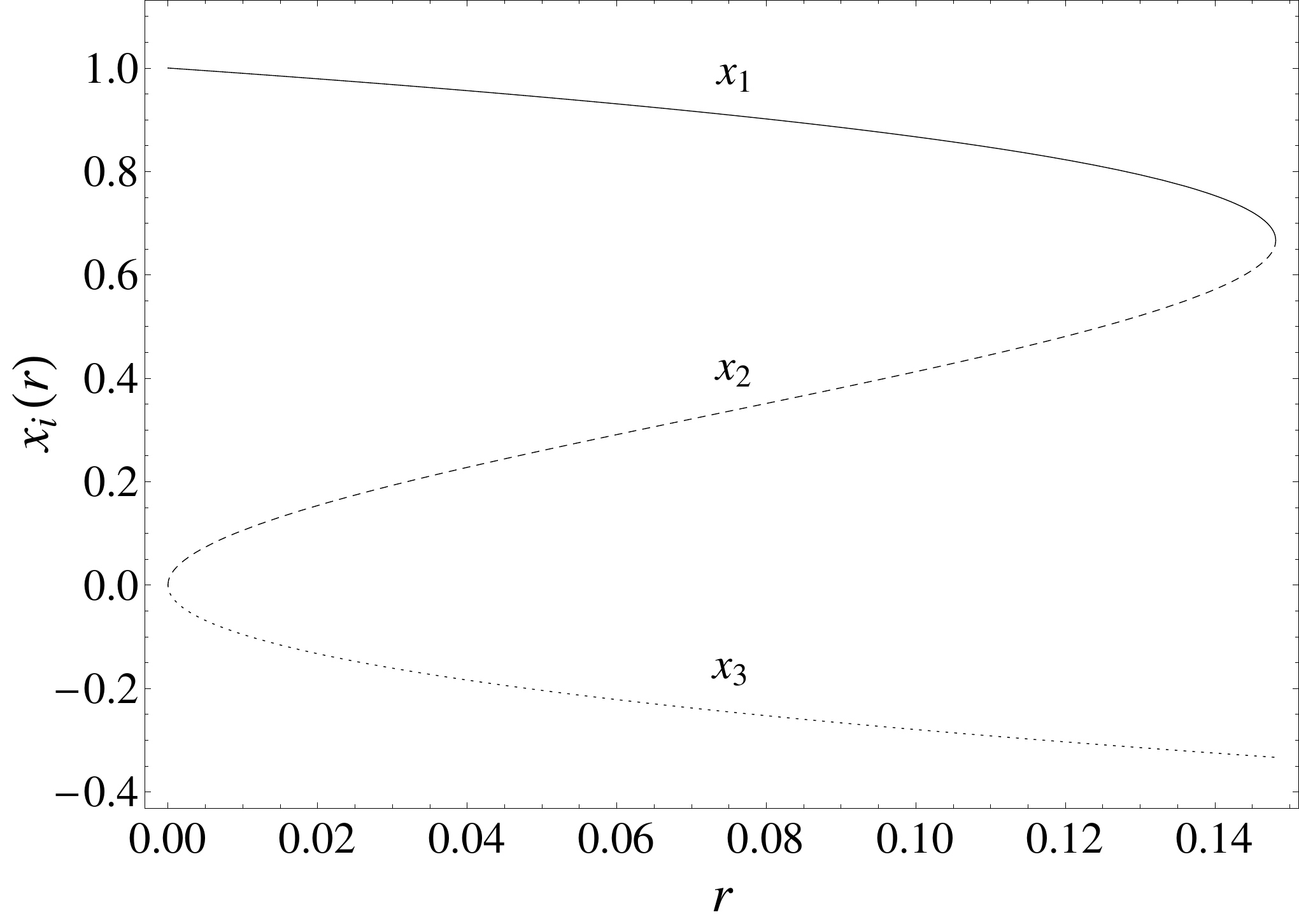}
\hspace{0.3cm}
\includegraphics[clip,width=0.48\textwidth]{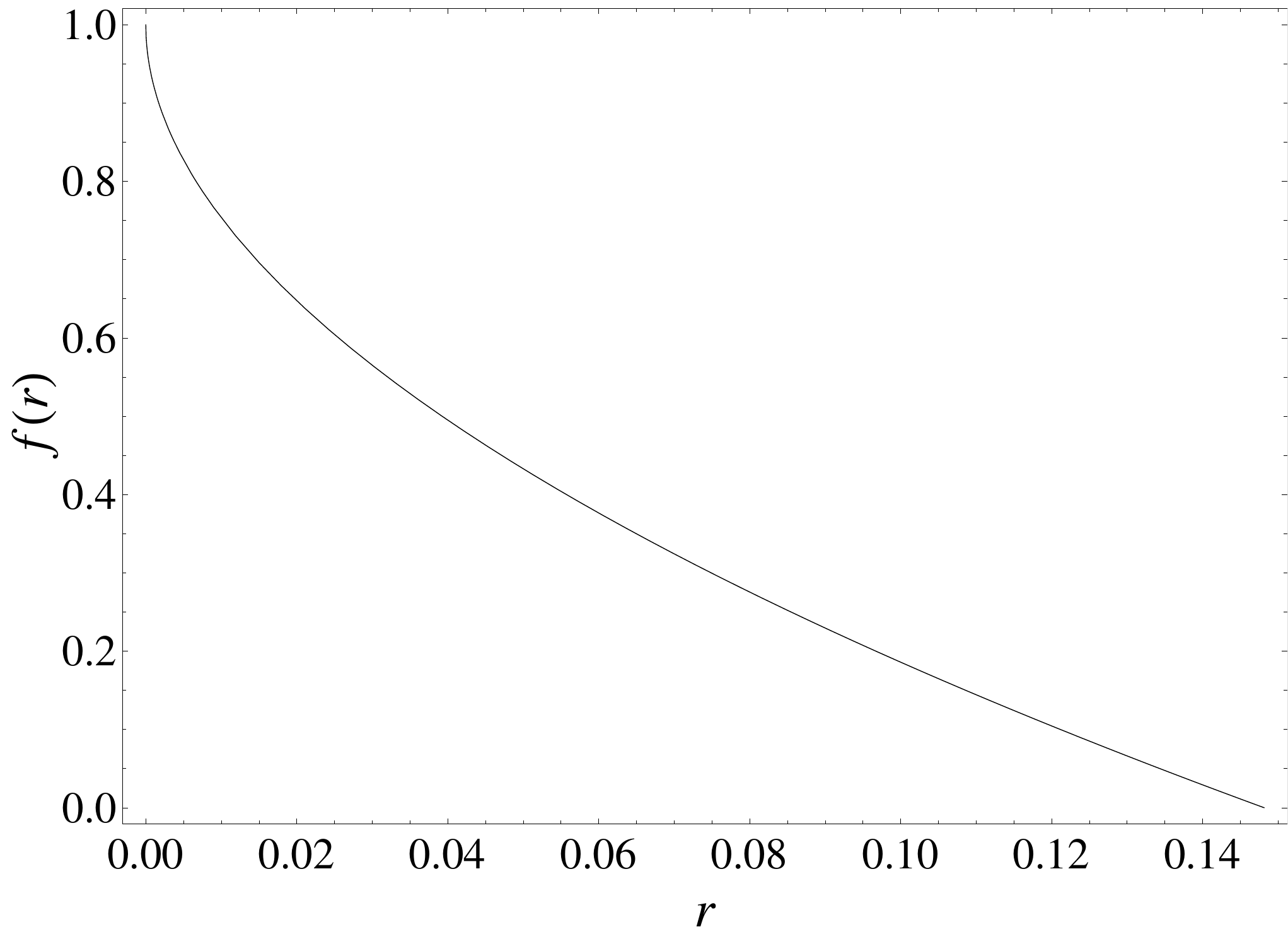}
\caption{{\it Left panel}. The functions $x_1(r)$, $x_2(r)$, and $x_3(r)$ in
Eqs.~(\ref{x1})-(\ref{x3}) for $0 \leq r \leq 4/27$.
Observe that $x_1(4/27) = x_2(4/27) = 2/3$ and $x_3(4/27) = -1/3$.
{\it Right panel}. The function $f(r)$ in Eq.~(\ref{f}) for $0 \leq r \leq 4/27$.}
\end{center}
\end{figure*}


{\it WKB approximation.} -- In this case, the WKB approximation is valid for
large values of $|p|$. The imaginary turning points are
\begin{equation}
\label{new1}
\alpha_I(p) = \frac{1}{6} \! \left[\ln (p^2/2V^2V_0) + i\pi (2n + 1)\right]\!, ~~n \in \mathbb{N}.
\end{equation}
%
Accordingly,
\begin{equation}
\label{new2}
\sigma_p = - \frac{q}{3} + \frac{|p|}{6} \ln \!\frac{|p|+ q}{|p|-q} \, ,
\end{equation}
%
where $q = \sqrt{-U_p(\alpha_R)} = \sqrt{p^2 + 2V^2V_0 e^{6\alpha_R}}$,
so that $\mbox{Im} \sigma_p = \pi |p|/6$. It follows that
\begin{equation}
\label{new3}
n_p = e^{-2\pi |p|/3},
\end{equation}
%
in agreement with Eq.~(\ref{HM}) in the case of large $|p|$.

{\it The case of null scalar potential.} -- In the case of flat universes with null
scalar potential, the in and out $\psi$ modes are normalized plane waves. As in the case
of conformally flat quantum theories in curved space, there is no production
of ``particles'' out from the vacuum. The number of created universes is then exactly zero.

\section{V. Creation of closed inflationary universes}

The problem does not admit an exact analytical solution.

\subsection{Va. Large universe number: small $|p|$}

{\it WKB approximation.} -- Let us work in WKB approximation and consider Eqs.~(\ref{sm2}) and (\ref{sm3}).
Using the change of variable $x = 2V_0 e^{2\alpha}$, the universe number can be written as
\begin{equation}
\label{npf}
n_p = e^{2\pi^2 f(r)/3V_0},
\end{equation}
where $r$ is given by Eq.~(\ref{r}) with $V = 2\pi^2$, and we have introduced the function
\begin{equation}
\label{fInt}
f(r) = \frac32 \int_{x_2}^{x_1} \frac{dx}{x} \sqrt{-x^3+x^2-r}.
\end{equation}
Here, $x_1 \geq x_2 \geq 0$ correspond to the classical turning points, the real and positive
solutions of the equation $x^3 - x^2 + r = 0$. The three solution of such a cubic equation
can be written as
\begin{eqnarray}
\label{x1}
&& x_1(r) = \frac13 \left( \! 1 + 2\cos \frac{\theta}{3} \right) \! , \\
\label{x2}
&& x_2(r) = \frac13 \left( \! 1 - 2\cos \frac{\theta + \pi}{3} \right)\! , \\
\label{x3}
&& x_3(r) = \frac13 \left( \! 1 - 2\cos \frac{\theta - \pi}{3} \right) \! ,
\end{eqnarray}
with
\begin{equation}
\label{theta}
\theta(r) = \arccos (1-27r/2).
\end{equation}
As it easy to check, the above three solutions are real when
\begin{equation}
\label{rcondition}
0 \leq r \leq 4/27.
\end{equation}
In this case, $x_1$ and $x_2$ are positive, with $x_1 \geq x_2$, and $x_3$ is negative (see
the left panel of Fig.~3).

The integral in Eqs.~(\ref{fInt}) can be expressed in terms of the complete elliptical integrals as
\begin{equation}
\label{f}
f(r) = c_K K (m) + c_E E (m) + c_{\Pi} \Pi (n,m),
\end{equation}
%
%
where
\begin{equation}
\label{ccc}
c_K = \frac{x_3}{\sqrt{x_2-x_3}} \, , ~~ c_E = \frac{x_2-x_3}{\sqrt{x_2-x_3}} \, , ~~
c_{\Pi} = \frac{3 x_1 x_3}{\sqrt{x_2-x_3}} \, ,
\end{equation}
and
\begin{equation}
\label{mn}
m = \frac{x_2-x_1}{x_2-x_3} \, , ~~ n = \frac{x_2-x_1}{x_2} \, .
\end{equation}
%
Here, $K(m)$, $E(m)$, and $\Pi(n,m)$ are the complete elliptical integrals of first,
second, and third kind, respectively~\cite{Abramowitz}.
A plot of the function $f(r)$ is shown in the right panel of Fig.~3.
Notice that
\begin{equation}
\label{lim}
\lim_{r \rightarrow 0}f(r) = 1, ~~ \lim_{r \rightarrow 4/27}f(r) = 0.
\end{equation}
The WKB result~(\ref{npf}) is valid only if $n_p \gg 1$, namely when the exponent
$2\pi^2 f(r)/3V_0$ is much bigger than unity. For $r \ll 1$ (the case $r \gg 1$ will
be analyzed in Section Vb), this means $V_0 \ll 2 \pi^2/3$. In this case, we have
\begin{equation}
\label{npclosedOK}
n_p = e^{2\pi^2/3V_0}, ~~ |p| \ll 1 \ll 2\pi^2/3V_0,
\end{equation}
for the average number of created universes.
The case $r \ll 1$ and $V_0 \gg 2 \pi^2/3$, namely $|p| \ll 2\pi^2/3V_0 \ll 1$,
cannot be solved in WKB approximation. We proceed as follows.

{\it Approximate Wheeler-DeWitt potential.} -- Let us approximate the Wheeler-DeWitt potential as
\begin{eqnarray}
\label{1c}
U_p(\alpha) \!\!& \simeq &\!\!
\left\{ \begin{array}{lll}
  -p^2 + 4\pi^4 e^{4\alpha},      & ~~~~ \alpha \leq \alpha_\star,
\\
  -p^2 - 8\pi^4 V_0 e^{6\alpha},  & ~~~~ \alpha > \alpha_\star,
  \end{array}
  \right.
\end{eqnarray}
where $e^{\alpha_\star} = 1/\sqrt{3V_0}$, and $\alpha_\star$ is the
point of maximum of the Wheeler-DeWitt potential. The approximate potential~(\ref{1c})
is discontinuous at $\alpha = \alpha_\star$ with a jump discontinuity of
\begin{equation}
\label{jump}
\Delta = |\!\! \lim_{\alpha \rightarrow \alpha_\star^{-}} \!\! U_p(\alpha) \,
- \lim_{\alpha \rightarrow \alpha_\star^{+}} \!\! U_p(\alpha)|
= \frac53 \! \left(\frac{2\pi^2}{3V_0} \right)^{\!2}.
\end{equation}
In the analogue case of quantum potential scattering,
the reflection and transmission coefficients obtained by approximating a smooth potential with
one possessing a jump discontinuity are trustworthy
only if the wavelength of the incident particle is much bigger than the square root of the
jump (see, e.g.~\cite{Campanelli}).
In our case, such a validity condition translates into the condition
\begin{equation}
\label{conditionDelta}
|p| \ll \sqrt{\Delta/2} \simeq 2\pi^2/3V_0.
\end{equation}
The Bunch-Davies-normalized $\psi_p^{({\rm in})}$ and $\psi_p^{({\rm out})}$ wave functions are easily found
in the case of the approximate Wheeler-DeWitt potential. They are
\begin{eqnarray}
\label{2c}
\!\!\!\!\!\!\!\!\!\!\! \psi_p^{({\rm in})} \!\!& = &\!\!
\left\{ \begin{array}{lll}
  u_p,                     & ~~~~ \alpha \leq \alpha_\star, \\
  c_1 v_p + c_2 v_p^*,     & ~~~~ \alpha > \alpha_\star,
  \end{array}
  \right.
\end{eqnarray}
and
\begin{eqnarray}
\label{3c}
\!\!\!\!\!\!\!\!\!\!\! \psi_p^{({\rm out})} \!\!& = &\!\!
\left\{ \begin{array}{lll}
   c_3 u_p + c_4 u_p^*,    & ~ \alpha \leq \alpha_\star, \\
   v_p,                    & ~ \alpha > \alpha_\star,
  \end{array}
  \right.
\end{eqnarray}
respectively. Here, $u_p$ is given by
\begin{equation}
\label{up}
u_p = \frac{\Gamma(1-ip/2)}{2^{ip/2} \sqrt{2p}} \, I_{-ip/2}(\pi^2e^{2\alpha}),
\end{equation}
where $\Gamma(x)$ is the Gamma function and $I_\nu(x)$ is the modified Bessel
function of first kind~\cite{Abramowitz}. The function $u_p$ represents a normalized
in-mode of a closed universe with $V_0 = 0$. The function $v_p$, instead, is given
by the right hand side of Eq.~(\ref{hp2}) and represent a normalized out-mode of a
flat universe with $V_0 \neq 0$.
The constants of integrations $c_i$ ($i=1,2,3,4$) can be found by imposing the
continuity of $\psi_p^{({\rm in})}$ and $\psi_p^{({\rm out})}$,
and their first derivatives, at $\alpha = \alpha_\star$. We find
\begin{eqnarray}
&& c_1 = c_3^*=
\langle \psi_p^{({\rm in})} | \psi_p^{({\rm out}) *} \rangle_{|\alpha=\alpha_\star}  = \alpha_p, \\
&& c_2 = -c_4 =
-\langle \psi_p^{({\rm in})} | \psi_p^{({\rm out})} \rangle_{|\alpha=\alpha_\star} = \beta_p.
\end{eqnarray}
Accordingly, the average number of universes is
\begin{equation}
\label{npclosed}
n_p = |\langle u_p | v_p \rangle|^2_{\alpha=\alpha_\star}.
\end{equation}
For $|p| \rightarrow 0$, or more precisely for $|p| \ll \min[1,2\pi^2/3V_0]$,
we find
\begin{equation}
\label{npclosed2}
n_p = \frac{H(V_0/2\pi^2)}{\pi |p|}
\end{equation}
at the leading order, where
\begin{eqnarray}
\label{hc}
&& \!\!\!  H(x) = \frac{\pi^2}{1296 x^2} \nonumber \\
&& \!\!\!  \times \!
\left[ \sqrt{6} I_1 \!\! \left(\frac{1}{6x}\right) \! H_0^{(1)}\!\! \left(\frac{\sqrt{6}}{27x} \right)
      + 2 I_0\!\! \left(\frac{1}{6x} \right) \! H_1^{(1)} \!\! \left(\frac{\sqrt{6}}{27x} \right) \right] \nonumber \\
&& \!\!\! \times \!
\left[ \sqrt{6} I_1 \!\! \left(\frac{1}{6x} \right) \! H_0^{(2)}\!\! \left(\frac{\sqrt{6}}{27x} \right)
      + 2 I_0 \!\! \left(\frac{1}{6x} \right) \! H_1^{(2)} \!\! \left(\frac{\sqrt{6}}{27x} \right)  \right] \!,
\nonumber \\
\end{eqnarray}
with $H_\nu^{(1)}(x)$ being the Hankel function of first kind~\cite{Abramowitz}.
Figure~4 shows the function $H(x)$ together with its asymptotic expansions
for small and large values of the argument,
\begin{eqnarray}
\label{hasyc}
H(x) \!\!& = &\!\!
\left\{ \begin{array}{lll}
   \frac{5}{4\sqrt{6}} \, e^{1/3x} \left( 1 + \mathcal{O}(x) \right) \! ,
                                                                   & ~ x \rightarrow 0,      \\
   3/2 + \mathcal{O}(1/x^2),                                       & ~ x \rightarrow \infty.
  \end{array}
  \right.
\end{eqnarray}
Accordingly, the average number of created universes for small $|p|$ is
\footnote{For large $|p|$, or more precisely for $|p| \gg \max[1,2\pi^2/3V_0]$,
Eq.~(\ref{npclosed}) would give an incorrect power-law decay for $n_p$,
instead of the correct exponential decay that will be derived in WKB approximation (see below).
This is due to the nonanalyticity of the approximate expression of the potential
$U_p(\alpha)$ at the point $\alpha_\star$. Indeed,
using perturbation theory and following~\cite{Landau} it is easy to find the expression
of the universe number in the case of large $|p|$. It turns out to be
\begin{equation}
\label{npclosed3}
n_p = \Delta^2/64 p^6 = 25\pi^8/729 V_0^4 p^4,
\end{equation}
where $\Delta$ is defined in Eq.~(\ref{jump}). We stress again that this result is
unphysical and follows from having approximated
the potential $U_p$ with a nonanalytical expression. Numerically, we checked that
Eq.~(\ref{npclosed}) ``correctly'' reduces to Eq.~(\ref{npclosed3}) for $|p| \gg \max[1,2\pi^2/3V_0]$.}
\begin{eqnarray}
\label{npskNEW}
n_p \!\!& \simeq &\!\!
\left\{ \begin{array}{lll}
   \frac{5}{4\sqrt{6} \, \pi |p|} \, e^{2\pi^2/3V_0},
   & ~ |p| \ll 1 \ll 2\pi^2/3V_0, \\ \\
   \frac{3}{2\pi |p|} \, ,
   & ~ |p| \ll 2\pi^2/3V_0 \ll 1. 
  \end{array}
  \right.
\end{eqnarray}
The first equation in~(\ref{npskNEW}) is
in agreement with the result~(\ref{npclosedOK}) obtained in WKB approximation.
Notice that both equations are approximate results and that, in general,
the WKB approximation cannot be used to calculate the pre-exponential factor
in the transmission coefficient~\cite{Landau} that, in our case,
corresponds to the reciprocal of the average number of created universes.

It is interesting to observe that for small $|p|$ and large values of the scalar potential,
$|p| \ll 2\pi^2/3V_0 \ll 1$, the number of closed universes approaches the number
of flat universes [see Eq.~~(\ref{HM})], and that the former is exponentially amplified
for small scalar potentials, $|p| \ll 1 \ll 2\pi^2/3V_0$.


\begin{figure}[t!]
\begin{center}
\includegraphics[clip,width=0.48\textwidth]{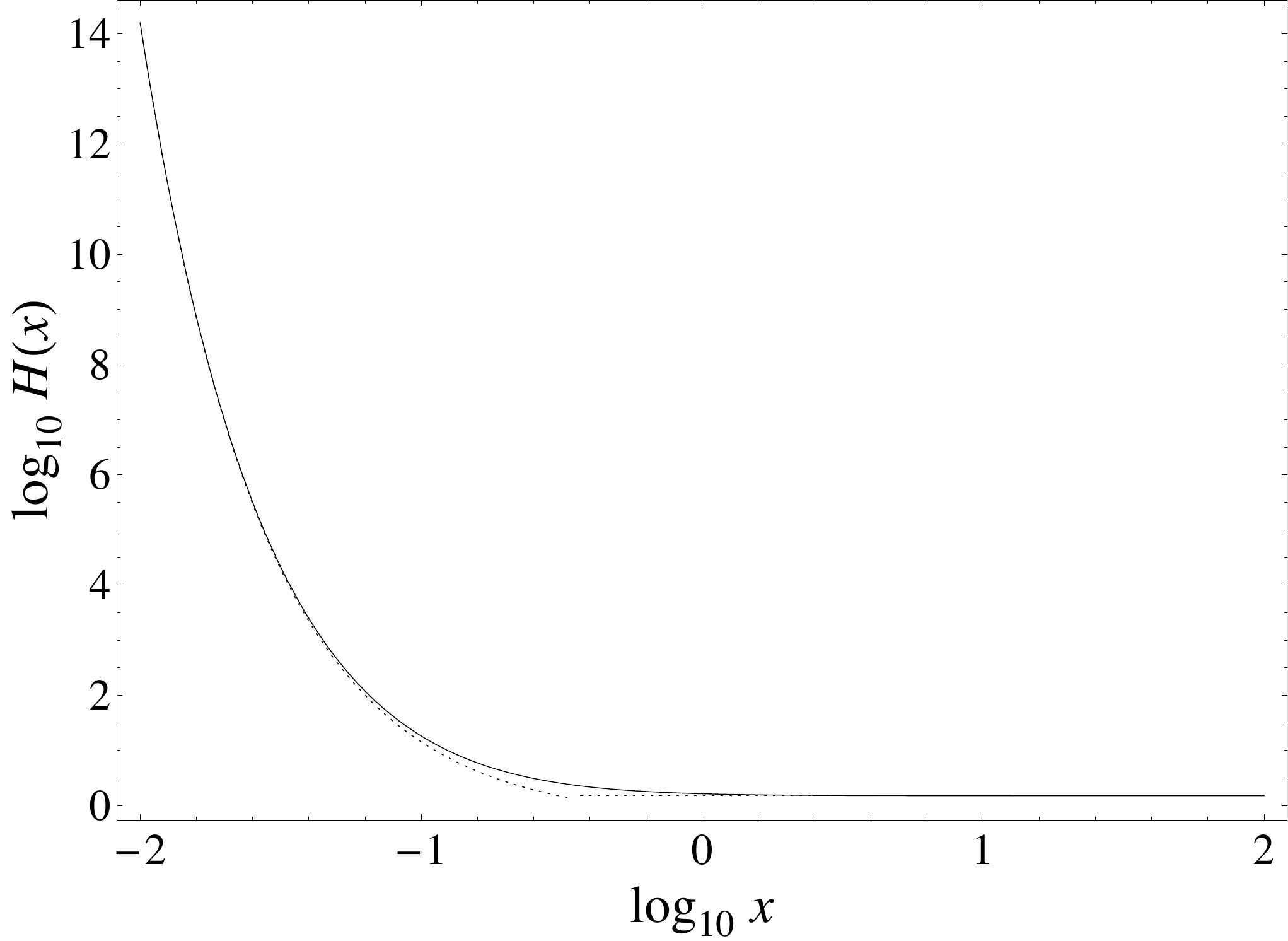}
\caption{The continuous and dotted lines represent, respectively,
the function $H(x)$ in Eq.~(\ref{hc}) and its asymptotic expansions in Eq.~(\ref{hasyc}).}
\end{center}
\end{figure}


\subsection{Vb. Small universe number: large $|p|$}

{\it WKB approximation.} -- The case of large $|p|$ can be only analyzed in WKB approximation (see footnote 2).
For $r > 4/27$, the solutions $x_1(r)$ and $x_2(3)$ are complex (conjugate), and $x_3(r)$ is negative. This
means that the Wheeler-DeWitt potential has no classical turning points. Using Eqs.~(\ref{sm4}) and (\ref{sm5}),
we find for the average number of created universes
\begin{equation}
\label{npg}
n_p = e^{-2\pi^2 g(r)/3V_0},
\end{equation}
where
\begin{equation}
\label{gInt}
g(r) = 3 \mbox{Im} \! \left[ \int_{x_R}^{x_1} \frac{dx}{x} \sqrt{x^3-x^2+r} \, \right] \!.
\end{equation}
Here, $x_{R}$ is a real and positive parameter, and between $x_1(r)$ and $x_2(r)$ we selected the former
as the imaginary turning point since it gives the smallest $\sigma_p$ (see discussion in Section IIIb).
Taking $x_{R} = -x_3(r)$, we find
\footnote{Interestingly enough, a numerical analysis shows that $g(r) = -f(r)$,
%
%
with $f(r)$ given by Eq.~(\ref{f}) for $r > 4/27$. We are not able to
provide an analytical proof of the above equality.}
\begin{equation}
\label{gg}
g(r) = -2 \mbox{Re} \! \left[ c_K F(\varphi,m) + c_E E(\varphi,m) + c_{\Pi} \Pi (n,\varphi,m) \right] \! ,
\end{equation}
%
%
where $c_K$, $c_E$, and $c_{\Pi}$ are given by Eq.~(\ref{ccc}), and
\begin{equation}
\label{varphi}
\varphi = \arcsin \! \sqrt{(x_2+x_3)/(x_2-x_1)} \, .
\end{equation}
%
Here, $F(\varphi,m)$, $E(\varphi,m)$, and $\Pi(n,\varphi,m)$
are the incomplete elliptical integrals of first, second, and third kind,
respectively~\cite{Abramowitz}.
Notice that
\begin{equation}
\label{limitg}
\lim_{r \rightarrow 4/27}g(r) = 0.
\end{equation}
A plot of $g(r)$ and its asymptotic expansion,
\begin{equation}
\label{gexp}
g(r) = \pi \sqrt{r} - C r^{1/6} + \mathcal{O}(r^{-1/6}),  ~~ r \rightarrow +\infty,
\end{equation}
is shown in Fig.~5. The constant $C$ in the above equation is defined by
\begin{equation}
\label{Cconstant}
C = \frac{3z_1 E(z_2) - \sqrt{3} z_1^*K(z_2)}{\sqrt{8}} \simeq 1.12025,
\end{equation}
%
where $z_1 = \sqrt{3 + i\sqrt{3}}$ and $z_2 = (1+i\sqrt{3})/2$.
Inserting the leading term of the asymptotic expansion~(\ref{gexp}) into Eq.~(\ref{npg}), we find
\begin{equation}
\label{npL}
n_p \simeq e^{-2\pi |p|/3}, ~~ |p| \gg 1,   
\end{equation}
for the average number of created universes. Thus, the
number of closed universes is exponentially suppressed for large $|p|$, as in the case
of flat universes [see Eq.~(\ref{HM})].


\begin{figure}[t!]
\begin{center}
\includegraphics[clip,width=0.48\textwidth]{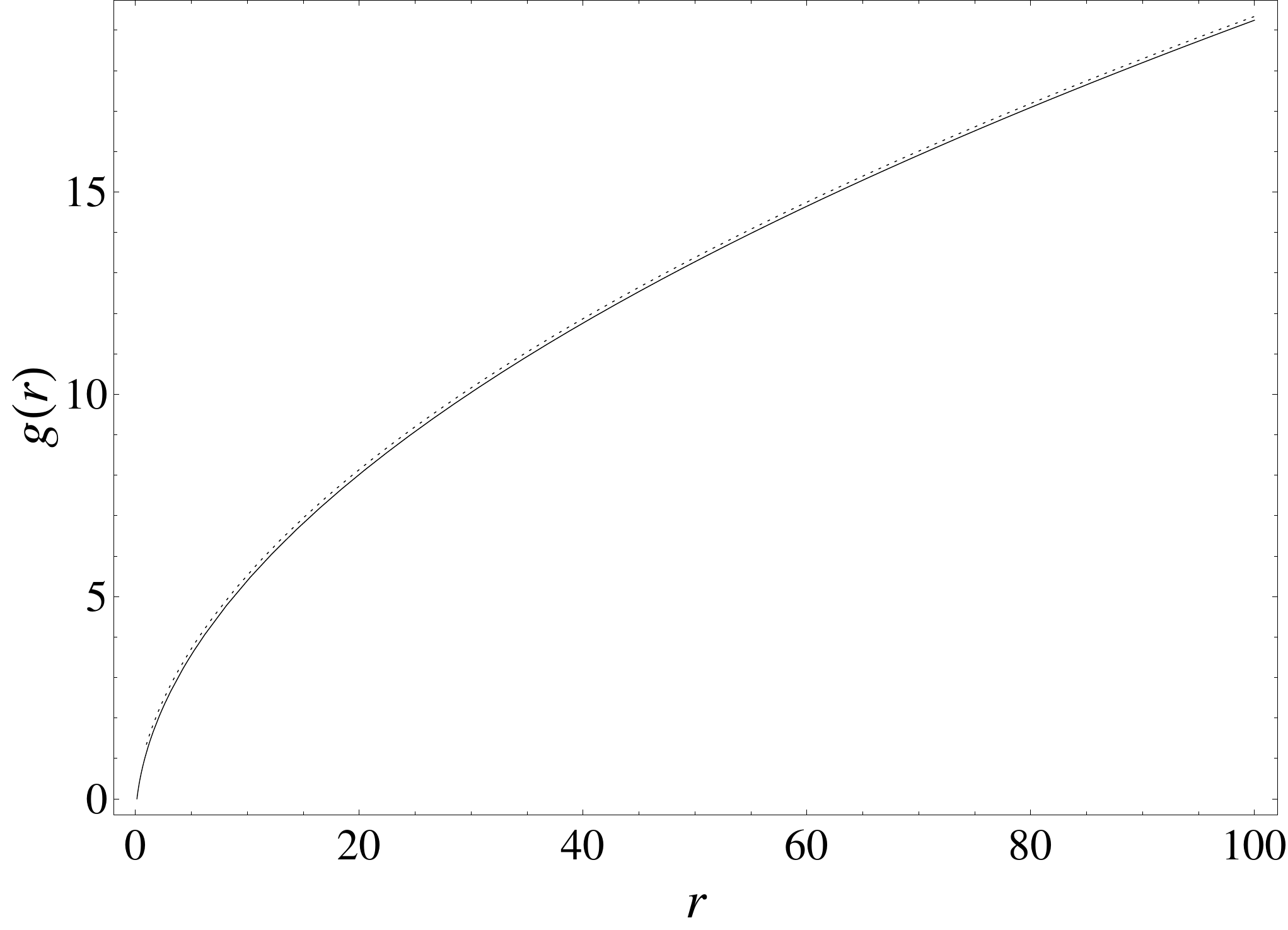}
\caption{The function $g(r)$ in Eq.~(\ref{gg}) for $r \geq 4/27$
with its asymptotic expansion~(\ref{gexp}).}
\end{center}
\end{figure}


{\it The case of null scalar potential.} -- In the case of closed universes with null
scalar potential, the out $\psi$ modes are exactly zero.
In this case, indeed, the Wheeler-DeWitt potential grows exponentially in the out region
($\alpha \rightarrow 0$), and represents an infinite potential barrier in the analogue case
of quantum potential scattering. No ``particles'' are present in the our region,
and the number of created universes is exactly zero.

The analogy between quantum potential scattering and the creation of universes from nothing
reposes on the assumption that the in and out modes are normalized according to the
Bunch-Davies ``prescription'', as in the case of quantum theory in curved space.
This in turns follows from the fact that the procedure of third quantization closely
mimics the one adopted in second quantization.
\footnote{Although in this paper we use the
{\it standard} Bunch-Davies vacuum, other possibilities cannot be excluded.
Indeed, the out-vacuum used by Kim~\cite{Kim} is not a Bunch-Davies normalized vacuum.
Not surprisingly, he found that the number of created closed universes in the case of
null scalar potential is different from zero. The number of created universes, thus,
strongly depends on the choice of the vacuum.
A similar situation occurs in second quantization, where the probability of creating a universe
strongly depends on the choice of the initial conditions for the wave
function of the Universe (see, e.g.,\cite{X}).}

\section{VI. Creation of open inflationary universes}

\subsection{VIa. The case of null scalar potential}

{\it Exact solution.} -- The case $k=-1$ and $V_0 = 0$ was analyzed by Kim~\cite{Kim}.
An exact solution for the number of created universes is given by
\begin{equation}
\label{Kim}
n_p = \frac{1}{e^{\pi |p|}-1}.
\end{equation}
For large $|p|$, $n_p $ is exponentially suppressed, while for small $|p|$, $n_p$ is inversely proportional to $|p|$.
Equation~(\ref{Kim}) is easily found by inserting the Bunch-Davies, in- and out-solutions
of Eq.~(\ref{motion3}) (with $k = -1$ and $V_0 = 0$),
\begin{eqnarray}
\label{jp}
&& \!\!\!\!\!\! \psi_p^{(\rm in)} = \sqrt{\frac{\pi}{4}} \,
                          \sinh^{-1/2}(p\pi/2) J_{-ip/2}(Ve^{2\alpha}/2),  \\
\label{jp2}
&& \!\!\!\!\!\! \psi_p^{(\rm out)} = \sqrt{\frac{\pi}{8}} \,
                          e^{-p\pi/4} H_{-ip/2}^{(2)}(Ve^{2\alpha}/2),
\end{eqnarray}
into Eq.~(\ref{alphabeta2}) and then using Eq.~(\ref{n}).

{\it WKB approximation.} -- In this case, the WKB approximation is valid for
large values of $|p|$. The imaginary turning points are
\begin{equation}
\label{nnew1}
\alpha_I(p) = \frac{1}{4} \left[\ln (p^2/V^2) + i\pi (2n + 1) \right] \!, ~~ n \in \mathbb{N}.
\end{equation}
%
Accordingly,
\begin{equation}
\label{nnew2}
\sigma_p = - \frac{s}{2} + \frac{|p|}{4} \ln \! \frac{|p|+ s}{|p|-s} \, ,
\end{equation}
%
where $s = \sqrt{-U_p(\alpha_R)} = \sqrt{p^2 + V^2e^{4\alpha_R}}$,
so that $\mbox{Im} \sigma_p = \pi |p|/4$. It follows that
\begin{equation}
\label{nnew3}
n_p = e^{-\pi |p|},
\end{equation}
in agreement with Eq.~(\ref{Kim}) in the case of large $|p|$.

\subsection{VIb. Small universe number: large $|p|$}


\begin{figure*}[t!]
\begin{center}
\includegraphics[clip,width=0.48\textwidth]{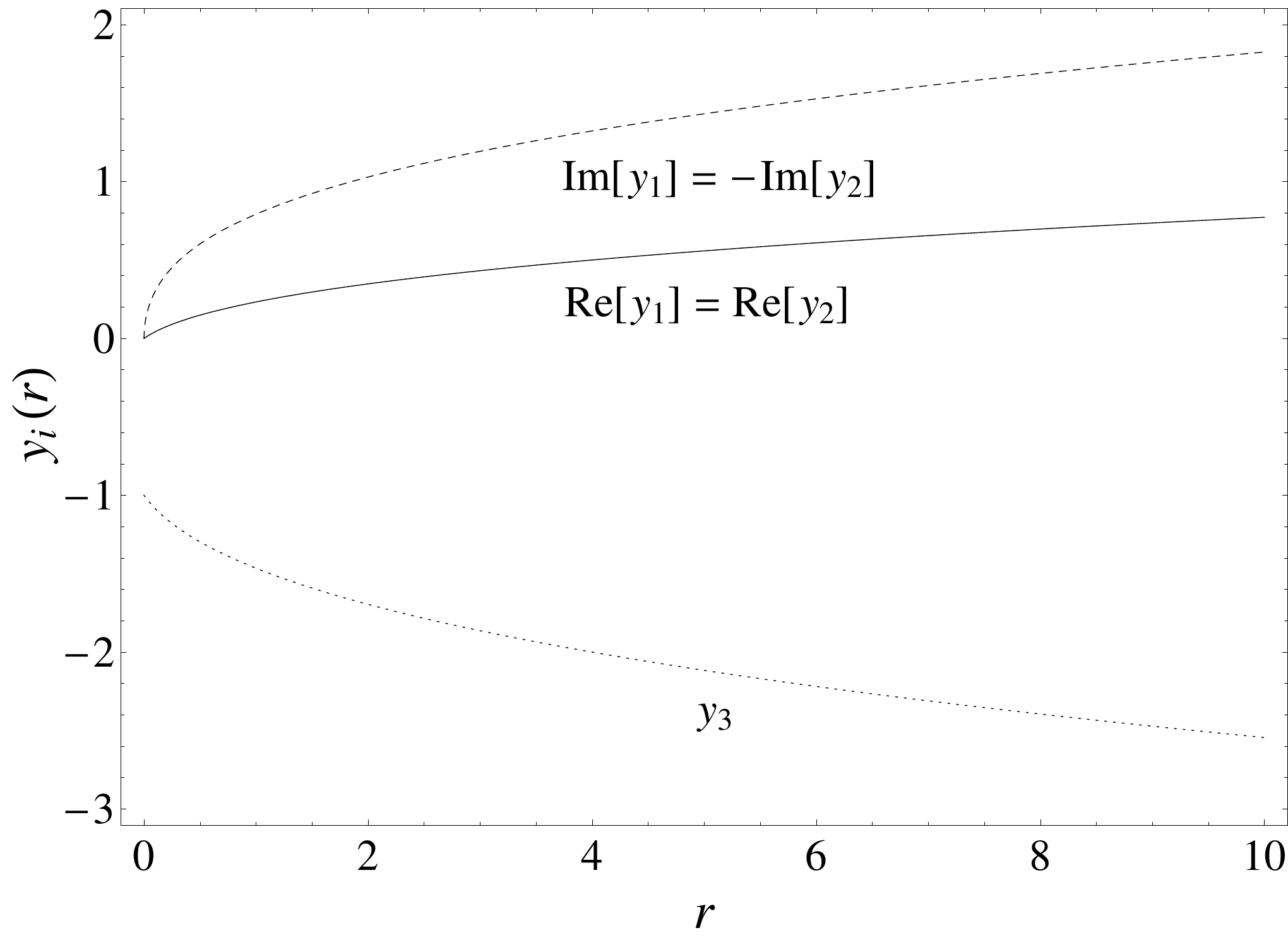}
\hspace{0.3cm}
\includegraphics[clip,width=0.48\textwidth]{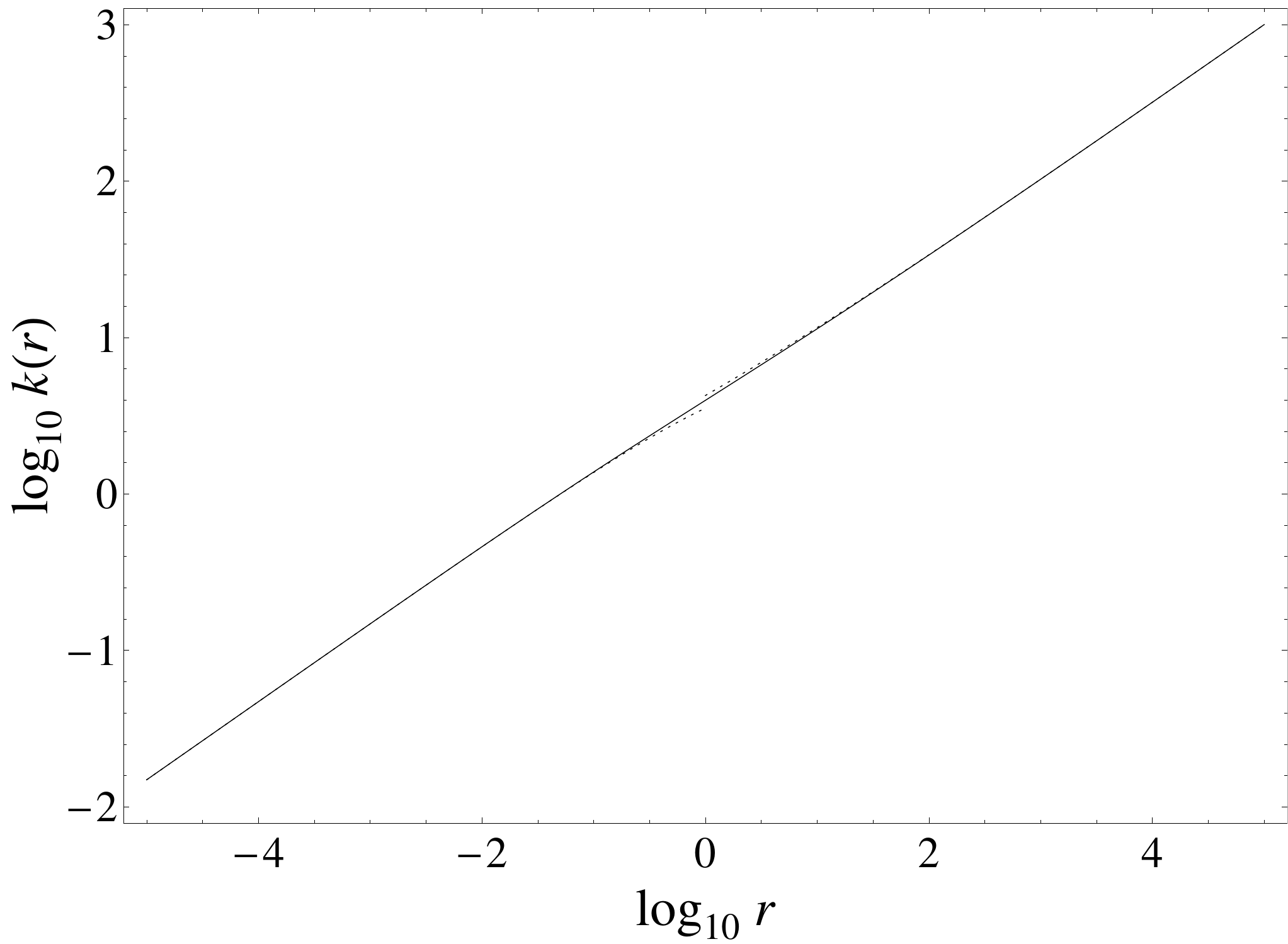}
\caption{{\it Left panel}. The functions $y_1(r)$, $y_2(r)$, and $y_3(r)$ in
Eqs.~(\ref{y1})-(\ref{y3}).
{\it Right panel}. The continuous and dotted lines represent, respectively,
the function $k(r)$ in Eq.~(\ref{kk}) and its asymptotic expansions in Eq.~(\ref{kasy}).}
\end{center}
\end{figure*}


{\it WKB approximation.} -- The case $k=-1$ and $V_0 \neq 0$ cannot be solved analytically.
Working in WKB approximation, we consider Eqs.~(\ref{sm4}) and (\ref{sm5}) since,
in this case, there are not classical turning points.
Using the change of variable $y = 2V_0 e^{2\alpha}$, the universe number can be written as
\begin{equation}
\label{npk}
n_p = e^{-V k(r)/3V_0},
\end{equation}
where
\begin{equation}
\label{kInt}
k(r) = 3 \mbox{Im} \! \left[ \int_{y_R}^{y_1} \frac{dy}{y} \sqrt{y^3+y^2+r} \, \right] \! .
\end{equation}
Here, $y_{R}$ is a real and positive parameter, while $y_1$ corresponds to the imaginary
turning point, the solution of the equation $y^3 + y^2 + r = 0$ which
gives the smallest $\sigma_p$ in Eq.~(\ref{sm5}).
The three solution of such a cubic equation can be written as
\begin{eqnarray}
\label{y1}
&& y_1 = -\frac13 \left(1 - 2\cos \frac{\vartheta}{3} \right) \! , \\
\label{y2}
&& y_2 = -\frac13 \left(1 + 2\cos \frac{\vartheta + \pi}{3} \right) \! , \\
\label{y3}
&& y_3 = -\frac13 \left(1 + 2\cos \frac{\vartheta - \pi}{3} \right) \! ,
\end{eqnarray}
with
\begin{equation}
\label{vartheta}
\vartheta(r) = \arccos (-1-27r/2).
\end{equation}
Notice that $\vartheta(r) = \pi - \theta(-r)$,
$y_1(r) = -x_3(-r)$, $y_2(r) = -x_2(-r)$, and $y_3(r) = -x_1(-r)$, where $\theta(r)$ is
defined in Eq.~(\ref{theta}), and $x_i(r)$ are given by Eqs.~(\ref{x1})-(\ref{x3}).
As it easy to check, $y_3$ is real and negative, while $y_1$ and $y_2$ are complex conjugate (see
the left panel of Fig.~6).
Taking $y_R = -y_3(r)$, the integral in Eqs.~(\ref{kInt}) can be expressed in terms of the
incomplete elliptical integrals as
\footnote{Numerically, we find that
$k(r) = C_K K(\mu) + C_E E(\mu) - C_{\Pi} \Pi (\nu,\mu)$. We are not able to
give an analytical proof of the above equality.}
\begin{equation}
\label{kk}
k(r) = 2\mbox{Re} \! \left[ C_K F(\omega,\mu)+ C_E E(\omega,\mu) - C_{\Pi} \Pi (\nu,\omega,\mu)\right] \!,
\end{equation}
where
\begin{equation}
\label{cccNew}
C_K = \frac{y_3}{\sqrt{y_2-y_3}} \, , ~~ C_E = \frac{y_2-y_3}{\sqrt{y_2-y_3}} \, , ~~
C_{\Pi} = \frac{3 y_1 y_3}{\sqrt{y_2-y_3}} \, ,
\end{equation}
\begin{equation}
\label{munu}
\mu = \frac{y_2-y_1}{y_2-y_3} \, , ~~ \nu = \frac{y_2-y_1}{y_2} \, ,
\end{equation}
and
\begin{equation}
\label{omega}
\omega = \arcsin \! \sqrt{(y_2+y_3)/(y_2-y_1)} \, .
\end{equation}
We show the graph of the function $k(r)$ in the
right panel of Fig.~6. Also shown are the asymptotic expansions of $k(r)$ for small and
large values of the argument,
\begin{eqnarray}
\label{kasy}
k(r) \!\!& = &\!\!
\left\{ \begin{array}{lll}
   \frac32 \pi \sqrt{r} -\frac38 \pi r + \mathcal{O}(r^2),  & ~ r \rightarrow 0,      \\
   \pi \sqrt{r} + C r^{1/6} + \mathcal{O}(r^{-1/6}),        & ~ r \rightarrow \infty,
  \end{array}
  \right.
\end{eqnarray}
where the constant $C$ is given by Eq.~(\ref{Cconstant}).
Inserting the leading terms of the above asymptotic expansions into Eq.~(\ref{npk}), we find
\begin{eqnarray}
\label{npLk}
n_p \!\!& \simeq &\!\!
\left\{ \begin{array}{lll}
   e^{-\pi |p|},     & ~ 1 \ll |p| \ll V/2V_0,  \\
   e^{-2\pi |p|/3},  & ~ |p| \gg \max[1,V/2V_0].
  \end{array}
  \right.
\end{eqnarray}
Thus, the number of open universes is exponentially suppressed for large $|p|$.
If the scalar potential is small, $V/2V_0 \gg 1$, the suppression factor is
the same as in the case of open universes with null scalar potential [see Eq.~(\ref{Kim})],
while if the scalar potential is large, $V/2V_0 \ll 1$, the suppression is similar
to that of flat universes [see Eq.~(\ref{HM})].

\subsection{VIc. Large universe number: small $|p|$}

{\it Approximate Wheeler-DeWitt potential.} -- The case of small $|p|$ cannot be solved in WKB approximation.
Let us proceed as in Section Va by approximating the Wheeler-DeWitt potential as
\begin{eqnarray}
\label{1}
U_p(\alpha) \!\!& \simeq &\!\!
\left\{ \begin{array}{lll}
  -p^2 - V^2 e^{4\alpha},       & ~~~~ \alpha \leq \alpha_*,
\\
  -p^2 - 2V^2 V_0 e^{6\alpha},  & ~~~~ \alpha > \alpha_*,
  \end{array}
  \right.
\end{eqnarray}
where $e^{\alpha_*} = 1/\sqrt{2V_0}$. The approximate potential is continuous at
$\alpha_*$, while its derivative is discontinuous with a jump discontinuity of
\begin{equation}
\label{jump2}
\delta = |\!\! \lim_{\alpha \rightarrow \alpha_\star^{-}} \!\! \dot{U}_p(\alpha) \,
- \lim_{\alpha \rightarrow \alpha_\star^{+}} \!\! \dot{U}_p(\alpha)|
= \frac{V^2}{2V_0^2} \, .
\end{equation}
As discussed in Section Va, the above approximation is trustworthy only
for values of $|p|$ small compared to the square root of the jump,
\begin{equation}
\label{conditionDelta}
|p| \ll \sqrt{|\alpha_*|\delta/2} \sim V/2V_0.
\end{equation}
The Bunch-Davies-normalized $\psi_p^{({\rm in})}$ and $\psi_p^{({\rm out})}$
wave functions are easily found in the case of the approximate Wheeler-DeWitt potential. They are
\begin{eqnarray}
\label{2}
\!\!\!\!\!\!\!\!\!\!\! \psi_p^{({\rm in})} \!\!& = &\!\!
\left\{ \begin{array}{lll}
  w_p,                     & ~~~~ \alpha \leq \alpha_*,
  \\
  c_1 v_p + c_2 v_p^*,     & ~~~~ \alpha > \alpha_*,
  \end{array}
  \right.
\end{eqnarray}
and
\begin{eqnarray}
\label{3}
\!\!\!\!\!\!\!\!\!\!\! \psi_p^{({\rm out})} \!\!& = &\!\!
\left\{ \begin{array}{lll}
   c_3 w_p + c_4 w_p^*,    & ~ \alpha \leq \alpha_*,  \\
   v_p,                    & ~ \alpha > \alpha_*,
  \end{array}
  \right.
\end{eqnarray}
respectively. Here, $w_p$ is given by the right hand side of Eq.~(\ref{jp}) and represents a
normalized in-mode of a open universe with $V_0 = 0$, while $v_p$ is given by the right hand side of Eq.~(\ref{hp2}) and represent a normalized out-mode of a flat universe with $V_0 \neq 0$.
The constants of integrations $c_i$ ($i=1,2,3,4$) can be found by imposing the continuity of $\psi_p^{({\rm in})}$ and $\psi_p^{({\rm out})}$,
and their first derivatives, at $\alpha_*$. We find
\begin{eqnarray}
&& c_1 = c_3^*=
\langle \psi_p^{({\rm in})} | \psi_p^{({\rm out}) *} \rangle_{|\alpha=\alpha_*}  = \alpha_p, \\
&& c_2 = -c_4 =
-\langle \psi_p^{({\rm in})} | \psi_p^{({\rm out})} \rangle_{|\alpha=\alpha_*} = \beta_p.
\end{eqnarray}
Accordingly, the average number of universes is
\begin{equation}
\label{npopen}
n_p = |\langle w_p | v_p \rangle|^2_{\alpha=\alpha_*}.
\end{equation}
For $|p| \rightarrow 0$, or more precisely for $|p| \ll \min[1,V/2V_0]$,
we find
\begin{equation}
\label{npopen2}
n_p = \frac{h(V_0/V)}{\pi |p|}
\end{equation}
at the leading order, where
\begin{eqnarray}
\label{h}
&& \!\!\!\!\!\!\!\!\! h(x) = \frac{\pi^2}{96 x^2} \nonumber \\
&& ~\, \times \! \left[ J_1(1/4x) H_0^{(1)}(1/6x) - J_0(1/4x) H_1^{(1)}(1/6x) \right] \nonumber \\
&& ~\, \times \! \left[ J_1(1/4x) H_0^{(2)}(1/6x) - J_0(1/4x) H_1^{(2)}(1/6x) \right] \!.
\nonumber \\
\end{eqnarray}
Figure~7 shows the function $h(x)$ together with its asymptotic expansions
for small and large values of the argument,
\begin{eqnarray}
\label{hasy}
h(x) \!\!& = &\!\!
\left\{ \begin{array}{lll}
   1 + x \cos (1/2x) + \mathcal{O}(x^2),  & ~ x \rightarrow 0,      \\
   3/2 + \mathcal{O}(1/x^2),              & ~ x \rightarrow \infty.
  \end{array}
  \right.
\end{eqnarray}
Therefore, at the leading order
\footnote{For large $|p|$, or more precisely for $|p| \gg \max[1,V/2V_0]$,
Eq.~(\ref{npopen}) would give an incorrect power-law decay for $n_p$,
instead of the correct exponential decay previously derived in WKB approximation. This,
as already discussed in footnote 4, is
due to the nonanalyticity of the potential $U_p(\alpha)$ at the point $\alpha_*$.
Using perturbation theory~\cite{Landau}, it is easy to find
\begin{equation}
\label{npopen3}
n_p = \delta^2/64 p^6 = V^4/256 V_0^4 p^6,
\end{equation}
where $\delta$ is defined by Eq.~(\ref{jump2}).
Numerically, we checked that
Eq.~(\ref{npopen}) ``correctly'' reduces to the ``unphysical'' result~(\ref{npopen3})
for $|p| \gg \max[1,V/2V_0]$.}
\begin{eqnarray}
\label{npsk}
n_p \!\!& \simeq &\!\!
\left\{ \begin{array}{lll}
   \frac{1}{\pi |p|},  & ~ |p| \ll 1 \ll V/2V_0, \\ \\
   \frac{3}{2\pi |p|}, & ~ |p| \ll V/2V_0 \ll 1. 
  \end{array}
  \right.
\end{eqnarray}
Thus, for small $|p|$ and small values of the scalar potential,
$|p| \ll 1 \ll V/2V_0$, the number of open universes approaches the number
of open universes with null scalar potential [see Eq.~(\ref{Kim})], while
for small $|p|$ and large values of the scalar potential,
$|p| \ll V/2V_0 \ll 1$, it approaches the number of flat universes [see Eq.~(\ref{HM})].


\begin{figure}[t!]
\begin{center}
\includegraphics[clip,width=0.48\textwidth]{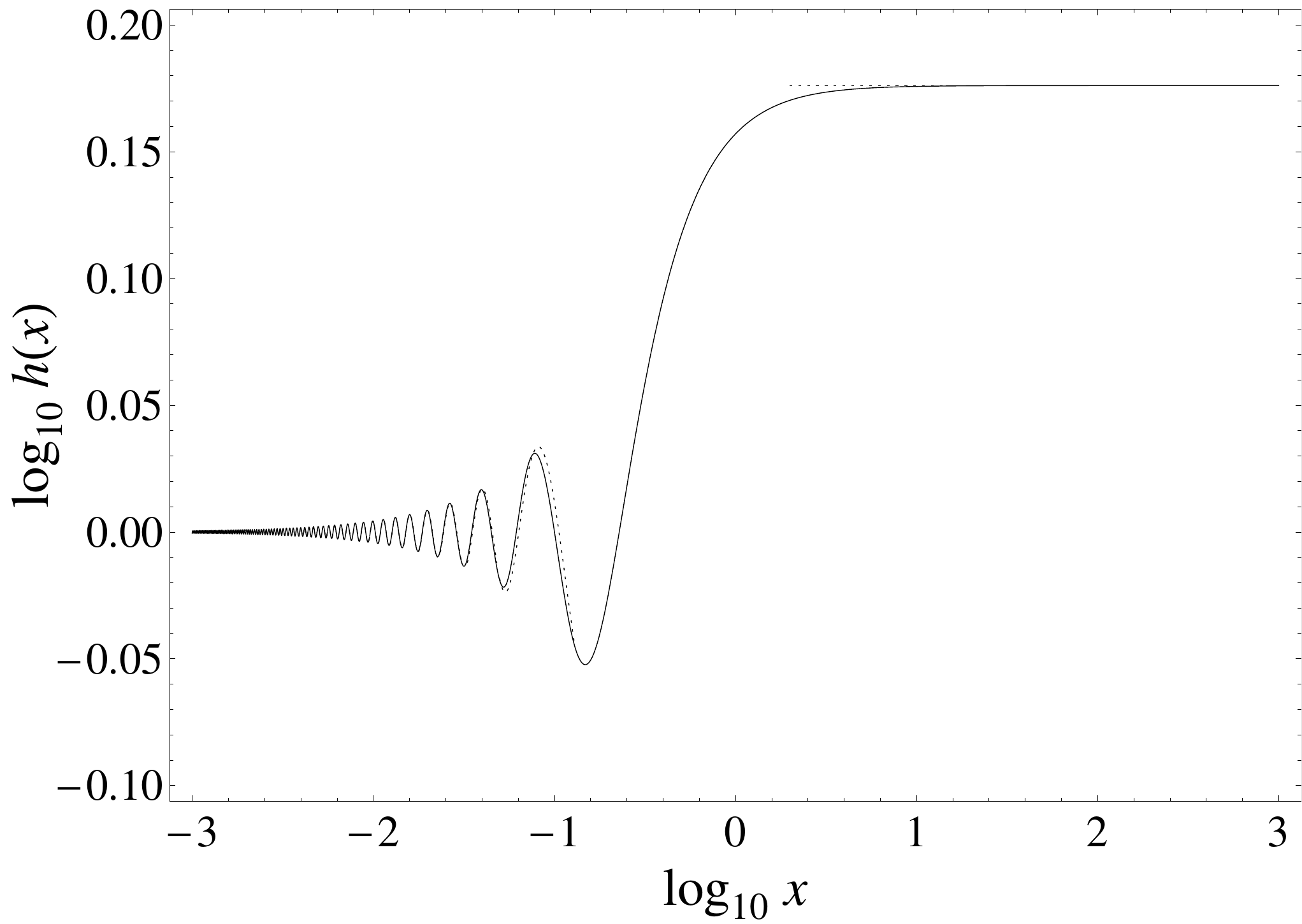}
\caption{The continuous and dotted lines represent, respectively,
the function $h(x)$ in Eq.~(\ref{h}) and its asymptotic expansions in Eq.~(\ref{hasy}).}
\end{center}
\end{figure}


\section{VII. Discussion and Conclusions}

{\it Discussion.} -- In order to be consistent with cosmic microwave background observations,
the scale of inflation $V_0^{1/4}$, which is directly
related to the amplitude of the primordial tensor perturbations, has to be below
$1.7 \times 10^{16}$GeV~\cite{Planck}.
The minimum value for the so-called ``reheat temperature'' is around $4.7 \,$MeV~\cite{MM}.
This constraint, which comes from the analysis of cosmic microwave background radiation data,
assumes a scale of inflation greater than about $43\,$MeV, which can be taken as a lower limit
for $V_0^{1/4}$. In the units used in this paper, these limits on the scale of
inflation translate into the constraint
\begin{equation}
\label{V0limit1}
2.7 \times 10^{-81} \lesssim V_0 \lesssim 6.6 \times 10^{-11}
\end{equation}
for the value of the scalar potential.
Since, $V_0 \ll 1$, the number of created universes from the third-quantized vacuum is
\begin{eqnarray}
\label{number}
n_p \!\!& \sim &\!\!
\left\{ \begin{array}{lll}
   \frac{1}{|p|} e^{2\pi^2/3V_0},    & ~ |p| \ll 1,  ~ (k = 1), \\
\\
  \frac{1}{|p|},                     & ~ |p| \ll 1, ~ (k = 0,-1), \\
\\
   e^{-c \pi |p|},                   & ~ |p| \gg 1, ~ (k = -1,0,1),
  \end{array}
  \right.
\end{eqnarray}
where $c = 2/3$ for closed and flat universes, and for open universes with $|p| \gg \max[1,V/V_0]$,
while $c = 1$ for open universes with $1 \ll |p| \ll V/V_0$.

Thus, universes with large values of $|p|$ are essentially not created, while the
creation from nothing occurs only for those universes labelled by small values of $|p|$.

Closed universes that are created in the out region with $a \gtrsim a_{\rm cr}$
undergo inflation since, in this case, $r = p^2V_0^2/\pi^4 \ll 1$ (see the middle panel of Fig.~2).
After creation, flat universes can either be kinetic-energy dominated or inflate.
Newly created open universes can inflate, be kinetic-energy dominated, or
curvature-dominated. For flat and open universes, the type of classical evolution after
creation depends of the the value of the parameter $r = 4p^2V_0^2/V^2$
and on the ``size'' $a$ of the created universe (see the upper and lower panel of Fig.~2,
respectively).

For small $|p|$ (namely for values of $|p|$ such that universes are effectively created),
the ratio of the number of closed universes to either the number of flat or open
universes is given by the factor $e^{2\pi^2/3V_0}$. Using Eq.~(\ref{V0limit1}),
this ratio is given by
%
\begin{equation}
\label{n01}
10^{10^{10}} \lesssim \frac{n_{closed}}{n_{flat,open}} \lesssim 10^{10^{81}}.
\end{equation}
Interestingly enough, recent analyses of the Planck data on the Cosmic Microwave Background radiation
favour a positive-curvature Universe~\cite{Silk,Handley}.

{\it Conclusions.} -- The creation from nothing of closed, open, and flat universes in the presence of
a scalar field (the inflaton) is a general consequence of third quantization.
Solving the Wheeler-DeWitt equation both in WKB approximation and using a suitable
approximation of the Wheeler-DeWitt potential, we have found that
the creation of universes, both closed or open and flat, is inhibited for universes with large
amounts of kinetic energy of the inflaton. For small values of the kinetic energy, instead,
closed, open, and flat universes are created from the third-quantized vacuum, the state of ``nothingness''.
Due to the relatively small value of the inflaton potential, as observed in our universe,
and for a given small amount of scalar kinetic energy, the creation of closed universes
is exponentially favoured over the creation of flat and open ones.






\end{document}